\documentstyle[prb,aps,twocolumn,floats,epsf]{revtex}
\begin{document}
\draft
\title{Quantum-Statistical Current Correlations in Multi-Lead Chaotic
Cavities}
\author{S. A. van Langen$^1$ and M. B{\"u}ttiker$^2$}
\address{$^1$ Instituut-Lorentz, Leiden University, P.O. Box 9506,
2300 RA Leiden, The Netherlands \\
$^2$ D\'epartement de physique th\'eorique,
Universit\'e de Gen\`eve, 24 Quai Ernest-Ansermet \\
CH-1211 Gen\`eve, Switzerland
\medskip\\
\parbox{14cm}{\rm
Quantum mechanics requires that identical particles are treated as 
indistinguishable. This requirement leads to correlations in
the fluctuating properties of a system. Theoretical predictions are made
for an experiment on a multi-lead chaotic quantum dot which can identify 
exchange effects in electronic current-current correlations. Interestingly,
we find that the ensemble averaged exchange effects are of the order
of the channel number, and are insensitive to dephasing.
}}
\maketitle

Quantum theory requires that identical particles are treated as
indistinguishable. In particular, the wave function has to remain
invariant up to a sign under the exchange of any pair of particles.
A well known consequence of this symmetry is the exchange hole in
the equal time density-density correlation of an electron gas. \cite{landau}
It is the purpose of this paper to investigate exchange 
effects in current-current correlations of a mesoscopic chaotic cavity
connected to four leads. A four-lead geometry is the simplest structure
which allows an unambiguous identification of exchange effects.
While a similar experiment has already been proposed and analyzed 
under conditions where electron motion is along one-dimensional 
edge channels, \cite{buttiker1} the investigation presented here is the
first for a non-trivial many-channel conductor. One might expect from the
former analysis that exchange effects are washed out in an ensemble of
chaotic conductors, and can thus be observed at best in the fluctuations
away from the average. The most important result of our discussion is that
this is not the case: exchange effects survive ensemble averaging and are
even of the same order of magnitude as the direct terms.

The general existence of exchange effects in a scattering process with
two detectors and two mutually incoherent sources has been pointed out
already by Goldberger et al., \cite{goldberger} following the
seminal experiments of Hanbury-Brown and Twiss \cite{hanbury} with
a stellar interferometer based on this principle. A clear account of
two-particle interference of bosons and fermions has been given by
Loudon. \cite{loudon}
In electronic conductors we deal with a Fermi sea, instead of only two
particles. Nevertheless, for electrons moving in a fixed Hartree potential
the current-current correlations can be expressed via two-particle exchange
amplitudes in terms of the single-particle scattering matrix. \cite{buttiker1}
The resulting correlations are negative, except in the case of 
normal-superconductor hybrid structures. \cite{anantram}
Our work is closely related to recent theoretical
\cite{dejong1,levitov} and experimental \cite{noiseexp} 
efforts to understand shot noise in mesoscopic conductors.
Most of this effort has concentrated on conductors which are effectively
two-terminal, and has focused on the suppression of the shot-noise power
below the uncorrelated Poisson limit $2e|I|$. The situation is
potentially richer in multi-lead conductors, where there is the possibility
of investigating correlations between current fluctuations at different leads.

We compute the current correlations for a chaotic quantum dot. \cite{blanter}
During the last few years, there has been an increasing activity of
experimentalists in tailoring electronic confinement potentials, and
characterizing the conductive properties of such cavities. \cite{chang,folk}
If the classical dynamics of the dot is fully chaotic, the quantum transport
properties are well described by a relatively simple statistical ensemble
for the scattering matrix. \cite{baranger,jalabert} Using this approach,
we find that both the direct and exchange contributions to the
correlations are of the order of the channel number $N$, and that they are
insensitive to dephasing.
This is remarkable, since in single-channel scattering geometries the
exchange terms depend sensitively on phases, in contrast with the direct
terms. \cite{buttiker1,loudon} Furthermore, we find that the sign
of the averaged exchange contribution reverses if the cavity is closed up
by tunnel barriers. The particular dependence of our results on the barrier
transparency offers a possibility to distinguish these correlations from 
other effects. In the end we compare with a purely classical resistor network.

The quantity that we study, is the zero-frequency spectral density of
current correlations,
\begin{equation}
\label{def}
P_{\alpha\beta}=2\int_{-\infty}^{\infty} \! dt \,
\overline{\Delta I_\alpha (t+t_0)\Delta I_\beta (t_0)},
\end{equation}
where $\Delta I_\alpha =I_\alpha -\overline{I_\alpha}$ is the fluctuation
of the current in lead $\alpha$ away from the time-average. It is shown in
Ref.\ \onlinecite{buttiker1} that
\begin{eqnarray}
\label{spectral1}
P_{\alpha\beta}= 2\frac{e^2}{h}\sum_{\gamma,\delta}\int \! dE \,
f_\gamma (1-f_\delta ) {\rm Tr}\left[
A_{\gamma\delta}(\alpha)A_{\delta\gamma}(\beta)\right]
\end{eqnarray}
where $f_\alpha (E)$ is the distribution of reservoir $\alpha$, and
\begin{eqnarray}
A_{\beta\gamma}(\alpha)=
1_\alpha\delta_{\alpha\beta}\delta_{\alpha\gamma}
-s^\dagger_{\alpha\beta}(E)s^{\phantom{\dagger}}_{\alpha\gamma}(E).
\end{eqnarray}
Here $s_{\alpha\beta}(E)$ is the sub-block of the scattering matrix $S$ for
scattering from lead $\beta$ ($N_\beta$ channels) to lead $\alpha$
at energy $E$, and $1_\alpha$ is the $N_\alpha\times N_\alpha$-unit matrix.
If all reservoirs are at zero-temperature equilibrium,
$f_\alpha (E)=\theta (E-eV_\alpha)$, the summation in
Eq.\ (\ref{spectral1}) is over $\gamma\ne\delta$, and the trace becomes
a noise conductance
\begin{equation}
\label{noisecond}
C_{\gamma\delta}(\alpha\beta )={\rm Tr}(
s^\dagger_{\alpha\gamma} s^{\phantom{\dagger}}_{\alpha\delta}
s^\dagger_{\beta\delta}  s^{\phantom{\dagger}}_{\beta\gamma}).
\end{equation}
To be specific we introduce the correlator $P_{34}$ for three experiments
A, B, and C: \cite{buttiker1}
In experiment A (B) a small voltage $V$ is applied only to lead 1 (2),
whereas in experiment C a voltage $V$ is applied to both reservoir 1 and 2.
It follows from Eq.\ (\ref{spectral1}) that the result of experiment C is
not identical to the sum of the results from experiments A and B.
We identify the difference as the exchange correlation:
\begin{equation}
\label{experiment}
P_{34}^{\rm ex} \equiv P_{34}^{\rm C} - P_{34}^{\rm A} - P_{34}^{\rm B} =
-2 P_0 {\rm Re}\, C_{12}(34),
\end{equation}
where $P_0=2e|V|e^2/h$. Although $P_{\alpha\beta}\leq 0$ in every experiment 
for $\alpha\neq\beta$, the difference can still have both signs.

\begin{figure}
\epsfxsize=0.6\hsize
\hspace*{\fill}
\epsffile{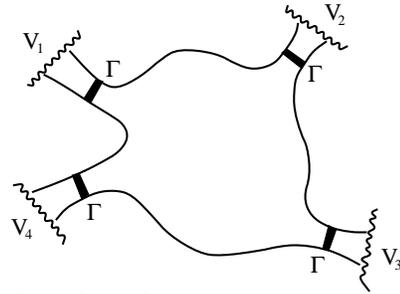}
\hspace*{\fill}
\caption{\label{fig1}
Chaotic cavity connected to four reservoirs via $N$-channel leads.
Tunnel barriers of transparency $\Gamma$ model non-ideal coupling
of the leads to the cavity.}
\end{figure}

We first consider ideal coupling of the four $N$-channel leads
to the cavity (Fig.\ \ref{fig1}). If time-reversal symmetry is (un)broken,
the $4N\times 4N$-scattering matrix $S$ is uniformly distributed on the
set of unitary (unitary symmetric) matrices. \cite{baranger,jalabert}
This is the circular unitary (orthogonal) ensemble of random matrix theory.
Since our results for both ensembles differ only to order $N^{-2}$, we will
focus on the simpler unitary ensemble, thus assuming that there is a
sufficiently large magnetic field in the cavity. From the general formula
of Ref.\ \onlinecite{creutz} one finds the exact ensemble average
$\langle\cdots\rangle$ of the noise conductances:
\begin{equation}
\label{ideal}
\langle C_{\gamma\delta}(\alpha\beta )\rangle =
(\delta_{\alpha\beta}+\delta_{\gamma\delta}-1/4) \times N^3/(16N^2-1).
\end{equation}
Fluctuations are of order one. Thus we find
\begin{equation}
\label{coruni}
\langle P_{\alpha\beta}\rangle= \frac{e^3}{h}
(\delta_{\alpha\beta}-1/4) \frac{N^3}{16N^2-1}
\sum_{i,j=1}^4 |V_i-V_j|.
\end{equation}
The average exchange correlation $\langle P_{34}^{\rm ex}\rangle =
\case{1}{2}P_0$ $N^3/(16N^2-1)$ is positive and, just as the direct terms,
of order $N$. This is remarkable, given the fact that the sign of the noise
conductance
$C_{12}(34)={\rm Tr}(s^\dagger_{31}s^{\phantom{\dagger}}_{32}
s^\dagger_{42} s^{\phantom{\dagger}}_{41})$ 
can vary from sample to sample. {\it The large average is essentially due to
correlations between scattering matrix elements imposed by unitarity.}

The size of the exchange effect, and the important role played by unitarity,
makes it more robust than a normal wave-interference effect. We next show
that the current correlations are the same even under conditions where
phase-coherent transfer through the sample is completely destroyed, but energy
is conserved in the dephasing process. We model dephasing by connecting
the sample to an additional fictitious reservoir. Attaching an equilibrium
reservoir, like a voltage probe, causes inelastic scattering, 
\cite{buttiker2} which reduces shot noise far below the value in 
phase-coherent transport, \cite{beenakker}
Following De Jong, \cite{dejong1,dejong2} we rather consider dephasing by a
reservoir with a fluctuating non-equilibrium distribution $f_\phi (E,t)$
such that no current is drawn {\it  at every energy and instant of time}.
Such quasi-elastic scattering does not change the shot noise of a chaotic
cavity with wide leads. \cite{dejong1} Below we generalize this result to
the current correlations in a multi-terminal geometry.

The total current through lead $\alpha$ at time $t$ and energy $E$
is \cite{buttiker1}
\begin{eqnarray}
\label{bu}
I_\alpha (E,t)=\frac{e}{h}\sum_\beta
(N_\alpha\delta_{\alpha\beta}- G_{\alpha\beta}) f_\beta
+\delta I_\alpha (E,t),
\end{eqnarray}
where $G_{\alpha\beta}={\rm Tr}\, s^\dagger_{\alpha\beta} 
s^{\phantom{\dagger}}_{\alpha\beta}$ and the indices run over all five leads.
Apart from the intrinsic fluctuations defined by $\delta I_\alpha (E,t)$,
there is now also a time-dependence due to the fluctuating $f_\phi (E,t)$.
The requirement $I_\phi (E,t)=0$ determines $f_\phi (E,t)$ which,
by substitution back into Eq.\ (\ref{bu}) yields the total fluctuation
$\Delta I_k$ at a real lead $k$, 
\begin{equation}
\label{total}
\Delta I_k = \delta I_k + G_{k\phi} \delta I_\phi /\sum_{m=1}^4 G_{\phi m}.
\end{equation}
The correlations of the intrinsic fluctuations
$\delta I_\alpha = \int\! dE\,\delta I_\alpha (E,t)$
satisfy Eq.\ (\ref{spectral1}) with the time-averaged distribution
\begin{equation}
f_\phi (E) =\sum_{k=1}^4 G_{\phi k}f(E-eV_k) / \sum_{k=1}^4 G_{\phi k}.
\end{equation}
We assume homogeneous and complete dephasing, which implies that the extra
lead has $N_\phi \gg N$ channels, and is coupled ideally to the
dot. \cite{brouwer1} In this limit, the only non-vanishing transport
coefficients are
\begin{mathletters}
\begin{eqnarray}
&& G_{k \phi}=G_{\phi k}=N,~~G_{\phi\phi}=N_\phi - 4N, \\
&& C_{kk}(\phi\phi) = N,~~C_{\phi\phi}(\phi\phi) = N_\phi - 4N,
\end{eqnarray}
\end{mathletters}%
independent of the magnetic field, for all $N$, and without fluctuations.
Thus
\begin{equation}
\label{dephas}
P_{kl}=2\frac{e^2}{h}(\delta_{kl}-1/4)N \int\! dE\, f_\phi (E) (1-f_\phi (E)).
\end{equation}
For $N\gg 1$ Eq.\ (\ref{dephas}) coincides exactly with Eq.\ (\ref{coruni}).
It shows a striking similarity with a semi-classical expression for the
shot-noise power first obtained by Nagaev. \cite{nagaev} However, it should
be noted that for geometries with non-ideal leads, discussed below, the
correlations are not determined by the distribution function of the
dephasing reservoir only.

We have now shown that exchange effects do survive changes in
the elastic scattering potential as well as phase-breaking scattering.
This finding sheds new light on the actual issue of shot-noise
suppression in two-terminal many-channel geometries. \cite{landauer}
Universal suppression factors were found in both quantum mechanical
\cite{beenakker,chen} and semi-classical \cite{dejong2,nagaev,davies}
approaches. In fact, the two-terminal shot noise contains direct and
exchange terms
corresponding to pairs of particles coming from different {\it channels}.
The reduction below the Poisson noise is partly due to such
channel-exchange terms.

The analysis of the cavity connected to open leads is not sufficient
for an unambigious identification of exchange correlations in practice.
Current correlations could be established by any other process, and there
is no general reason why the results of experiments A, B, and C should be
additive. Indeed, in the end we will briefly discuss a classical 
network where currents are correlated by fluctuations of the 
self-consistent electrostatic potential inside the ``dot'',
and where experiment C is not the sum of experiment A and B. 
To facilitate identification of the exchange effect, we now make
specific predictions for the correlations in the case of non-ideal
leads, as a function of the probability $\Gamma$ of transmission through
the contact region. We focus on the many-channel limit $N\gg 1$, where one
cannot distinguish isolated resonances of the system, except if the coupling
is so weak that $\Gamma \ll 1/N$. Below, we discuss this weak coupling
limit before turning to the main result of the paper: the correlations in
the regime $1/N \ll \Gamma \le 1$.

We first assume that the contacts are so poorly transmitting that transport
is dominated by a single state at the Fermi energy $E_F$.
The scattering amplitude from a point $b$ to a point $a$ takes the form
\begin{equation}
\label{sreso}
S_{ab}=\delta_{ab}-i\alpha \Delta/\pi
\frac{\psi_\nu^\ast(a)\psi_\nu(b)}{E_F-E_\nu+i\gamma_\nu/2}.
\end{equation}
where $\psi_\nu$ is the resonant eigenstate with energy $E_\nu$, $\Delta$
is the mean level spacing, $\alpha \ll 1$ is a dimensionless coupling
parameter, and $\gamma_\nu=\alpha \Delta/\pi\sum_r |\psi_\nu (r)|^2$ is the
width of the resonant level. The sum is over positions in the four coupling
regions. We assume that each lead contributes equally to the width,
which is automatically satisfied for a chaotic cavity with leads of the same
width, due to self-averaging of the overlaps with the eigenstate.
Then, at small bias $V\ll\gamma_\nu$, the current correlations are
\begin{mathletters}
\label{reso}
\begin{eqnarray}
&&P_{\alpha\beta}^{\rm A,B}=P_0 (4\delta_{\alpha\beta}-1)/16, \\
&&P_{\alpha\alpha}^{\rm C} = -P_{34}^{\rm C}= -P_{12}^{\rm C}= P_0 /4, \\
&&P_{13}^{\rm C}=P_{14}^{\rm C}=P_{23}^{\rm C}=P_{24}^{\rm C}=0.
\end{eqnarray}
\end{mathletters}%
The result of experiment C is easily understood. For resonant tunneling
through a two-terminal symmetric barrier, the shot noise vanishes.
Indeed, in experiment C the total current from lead 1 and 2 to lead 3 and 4
is noiseless. All fluctuations and
correlations are due to `unification' of the currents from 1 and 2,
and `partition' of the total current into 3 and 4. The vanishing of the
correlations between incoming and outgoing currents shows that these two
choices are made independently. The exchange correlation
$P_{34}^{\rm ex}=-P_0/8$ has a sign opposite to that of the ideally coupled
cavity.

We now compute the ensemble averaged correlations for the cavity with
tunnel barriers in the regime $N\gg 1$, $N\Gamma \gg 1$. The scattering
matrix of the combined system is \cite{brouwer2}
\begin{equation}
S = R + T'(1-UR')^{-1}UT,
\end{equation}
where we assume without loss of generality that the reflection and
transmission matrices of the barriers are proportional to the
$4N\times 4N$ unit matrix $I$: $R=R'=(1-\Gamma)^{1/2}I$, $T=-T'=\Gamma^{1/2}I$.
The distribution of the scattering matrix $U$ of the cavity is the circular
unitary ensemble, as before. The average noise conductances are computed
by series expansion of the four fractions $(1-UR')^{-1}$, collecting the
order $N$ contributions to each term with the diagrammatic technique
of Ref.\ \onlinecite{brouwer2}, and resumming. Thus one finds
\begin{eqnarray}
\label{gamma}
&&\langle C_{\gamma\delta}(\alpha\beta )\rangle = 
\left[\Gamma (2-3\Gamma )+4\Gamma^2 (\delta_{\alpha\beta}+\delta_{\gamma\delta}
)\right. \nonumber \\
&& - 4\Gamma(1-\Gamma )
(\delta_{\alpha\gamma}+\delta_{\alpha\delta}+\delta_{\beta\gamma}+
\delta_{\beta\delta}) \nonumber \\
&& + 16\Gamma (1-\Gamma)(
\delta_{\alpha\beta}\delta_{\beta\gamma} +
\delta_{\alpha\beta}\delta_{\beta\delta} +
\delta_{\alpha\gamma}\delta_{\gamma\delta} +
\delta_{\beta\gamma}\delta_{\gamma\delta}) \nonumber \\
&& \left. + 64 (1-\Gamma )^2
\delta_{\alpha\beta}\delta_{\beta\gamma}\delta_{\gamma\delta}
\right] \times N/64 +{\cal O}(1),
\end{eqnarray}
which for $\Gamma=1$ reduces to the large-$N$ limit of Eq.\ (\ref{ideal}).
The correlators for the experiments A and C are
\begin{mathletters}
\label{closedcorr}
\begin{eqnarray}
&&\langle P_{\alpha\beta}^{\rm C} \rangle =
P_0 N \Gamma (2-\Gamma )(4\delta_{\alpha\beta}-1)/16, \\
&&\langle P_{\alpha 1}^{\rm A}\rangle =
P_0 N\Gamma (10-7\Gamma )(4\delta_{\alpha 1}-1)/64, \\
&&\langle P_{\alpha\alpha}^{\rm A} \rangle 
=P_0N\Gamma (14-5\Gamma)/64~~(\alpha\neq 1),\\
&&\langle P_{\alpha\beta}^{\rm A} \rangle 
=-P_0N\Gamma (\Gamma +2)/64
~~(\alpha,\beta\neq 1, \alpha\neq \beta).
\end{eqnarray}
\end{mathletters}%
The functional dependence of these results provides a fingerprint 
for an experimental identification of the correlations. The exchange 
correlation
$\langle P_{34}^{\rm ex}\rangle = P_0 N\Gamma (3\Gamma -2)/32$
reverses sign at $\Gamma =2/3$.
An observation of this sign change would be a clear indication that measured
correlations are due to exchange. 
One easily checks that in experiment C the shot noise of the total current
from lead 1 and 2 to lead 3 and 4 
crosses over from one quarter times the Poisson noise at $\Gamma =1$ to one
half of the Poisson noise if $\Gamma$ is small, both in correspondence with
the literature. \cite{jalabert,chen,davies}

\begin{figure}
\epsfxsize=0.6\hsize
\hspace*{\fill}
\epsffile{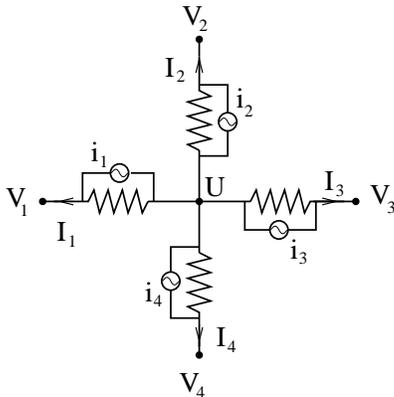}
\hspace*{\fill}
\caption{
\label{fig2}
Classical network consisting of resistors and noise sources.
The outgoing currents are correlated, due to fluctuations of the
``dot'' potential preventing temporal accumulations of charge.
}
\end{figure}

We now compare these results with a classical circuit, where
current conservation induces correlations, which are also non-additive
for the experiments A, B, and C. Four resistors $R$ are connected to
four voltage sources as in Fig.\ \ref{fig2}. Parallel to the resistors there are
independent sources of Poisson noise.
Temporal fluctuations of the central potential around $U=\sum_k V_k/4$ are
necessary to conserve the instantaneous total current. These voltage
fluctuations yield the following current correlations:
\begin{equation}
\label{resistors}
P_{kl}=e\sum_{m=1}^4
\left[1+4(2\delta_{kl}-1)(\delta_{mk}+\delta_{ml})\right]\frac{|V_m-U|}{8R}.
\end{equation}
For experiment A, B and C, Eq.\ (\ref{resistors}) gives the same result
as Eq.\ (\ref{closedcorr}) with $\Gamma\ll 1$ and $N\Gamma\, e^2/h=1/R$. For
other values of the voltages and $\Gamma$ the correlations differ.
This example demonstrates the need to investigate the correlations as a
function of a parameter like the barrier strength $\Gamma$ in order to
understand the source of the correlations. 

In conclusion, we have evaluated the exchange contributions to
current-current correlations in a multilead chaotic cavity.
We found that these correlations, instead of being a small interference
effect, are of the order of the channel number $N$, and that they
persist in the presence of dephasing. We have made specific predictions
for the dependence of the correlations on the transparency of non-ideal
leads. Finding such a direct signature of exchange in experiments is likely
a challenging task, but would clearly be a fundamental contribution to our
understanding of noise in electrical conductors.

Discussions with Ya.\ Blanter have helped us to sharpen our views.

S.\ v.\ L.\ was supported by the ``Ne\-der\-land\-se or\-ga\-ni\-sa\-tie voor
We\-ten\-schap\-pe\-lijk On\-der\-zoek'' (NWO) and by the ``Stich\-ting voor
Fun\-da\-men\-teel On\-der\-zoek der Ma\-te\-rie'' (FOM). 
M.\ B.\ was supported by the Swiss National Science Foundation.
\narrowtext

\end{document}